\documentclass[journal=jacsat,manuscript=article]{achemso}

\usepackage[version=3]{mhchem} 
\usepackage{booktabs}
\usepackage{cleveref}
\usepackage{graphicx}
\usepackage[T1]{fontenc}
\usepackage[utf8]{inputenc}
\usepackage{mathtools}
\usepackage{multirow}
\usepackage{tabularx}
\usepackage[section]{placeins}
\usepackage{array}
\usepackage{xcolor}

\author{Onofrio Tau}
\email{otau@icmab.es}
\affiliation[ICMAB]
{Institut de Ciència de Materials de Barcelona, ICMAB-CSIC, Campus UAB, 08193 Bellaterra, Spain}

\author{Giacomo Giorgi}
\affiliation[University of Perugia]
{Department of Civil and Environmental Engineering (DICA), Università degli Studi di Perugia, Via G. Duranti 93, 06125, Perugia, Italy}
\alsoaffiliation[CNR]
{CNR-SCITEC, I-06123, Perugia, Italy}
\alsoaffiliation[Ciriaf]
{CIRIAF - Interuniversity Research Centre, University of Perugia, Perugia, Italy}
\alsoaffiliation[CNR2]
{Centro S3, CNR-Istituto Nanoscienze, Via G. Campi 213/a, Modena, 41125, Italy, Italy}

\author{Riccardo Rurali}
\email{rrurali@icmab.es}
\affiliation[ICMAB]
{Institut de Ciència de Materials de Barcelona, ICMAB-CSIC, Campus UAB, 08193 Bellaterra, Spain}

\title[An \textsf{achemso} demo]
{CO$_2$ adsorption and photocatalytic reduction mechanisms on Ti-terminated CaTiO$_3$ (100) surface: a DFT study}

\abbreviations{n-CP. CP, PBFDO, TE}

\begin{document}

\begin{tocentry}

Some journals require a graphical entry for the Table of Contents.
This should be laid out ``print ready'' so that the sizing of the
text is correct.

Inside the \texttt{tocentry} environment, the font used is Helvetica
8\,pt, as required by \emph{Journal of the American Chemical
Society}.

The surrounding frame is 9\,cm by 3.5\,cm, which is the maximum
permitted for  \emph{Journal of the American Chemical Society}
graphical table of content entries. The box will not resize if the
content is too big: instead it will overflow the edge of the box.

This box and the associated title will always be printed on a
separate page at the end of the document.

\end{tocentry}

\begin{abstract}
  Photoreduction of CO$_2$ is an important alternative approach aimed to reduce the CO$_2$ atmospheric content which is responsible of the global warming. The development of an efficient photocatalyst can strongly improve the efficiency and selectivity of the by-products of such a process. Recently, CaTiO$_3$ has been used as an alternative semiconductor catalyst due to its attractive properties. In this study, we use calculations of the electronic structure of first principles to investigate for the first time the general reaction mechanism that leads to the main value-added by-products of HCOOH, CO, H$_3$COH and CH$_4$ byproducts, focusing on the reactions of adsorption, activation and reduction reactions of CO$_2$ molecules on the Ti-terminated CaTiO$_3$ (100) surface. We compute adsorption energies of the various intermediate configurations and activation energy barriers of the chemical reaction pathways. 
  Our results show that CO$_2$ can be activated by charge transfer of excess electrons leading to a CO$_2$ anion that probably gives HCOO by the first reduction; however, the second hydrogenation to HCOOH is impeded by the prohibitive energy barrier. In particular, activated CO$_2$ can also easily undergo decomposition, which facilitates CO production. Afterwards, we discuss the possible reaction mechanisms of CO photoreduction towards H$_3$COH and CH$_4$ value-added products, taking into account the experimental evidence that only CO and CH$_4$ have been detected. The reaction pathway generally follows the most energetically convenient routes characterized by activated intermediates. Even though H$_3$COH could be finally produced, its strong adsorption ($E_{ads}$ of -0.93 eV) and its promoted decomposition to H$_3$CO+H on the surface could explain why it has not been detected, as opposed to CH$_4$ whose $E_{ads}$ is only -0.22 eV due to its non-polar nature.
\end{abstract}


\section{1. Introduction}

The greenhouse effect has been responsible for the global warming, mainly due to the massive release of carbon dioxide (CO$_2$) by human activities which is recognized as one major greenhouse gas~\cite{low2017surface, habisreutinger2013photocatalytic}. To alleviate the atmospheric content of CO$_2$ and avoid severe environmental problems, CO$_2$ conversion in value-added chemicals by photocatalysis\cite{fu2024artificial, mori2024co2, yan20242d, gonzalez2021carbon, chen2019solar} has been proposed and gained increasing attention during the last decades. Inspired by natural photosynthesis, CO$_2$ is consumed by catalytic hydrogenation using solar light and semiconductor photocatalysts to enable the production of energy-rich chemicals like methane (CH$_4$), methanol (CH$_3$OH), formic acid (HCOOH) and other hydrocarbons that can be exploited as chemical feedstock and fuels\cite{kumar2012photochemical}. Moreover, the process can be considered environmentally friendly as it is also typically carried out at ambient reaction conditions since other transformation methods involve high process operational cost because of high pressure and temperature\cite{roy2010toward, tahir2013recycling, tu2014photocatalytic}. 
Although important progresses have been made since the first report\cite{inoue1979photoelectrocatalytic}, the photocatalytic reduction of CO$_2$ is still far away from industrial applications due to the low catalytic conversion and poor product selectivity\cite{fu2024artificial, li2014design}. This can be explained by two main following reasons: (i) CO$_2$ is highly themodynamically and kinetically stable due to its electronic structure; (ii) the CO$_2$ photoreduction to other products can be encouraged by activation, namely by bending the molecule in its initial linear shape on the catalytic surface, creating dipole moments and thus enhancing the chemical reactivity, but this typically requires a very high energy cost\cite{wang2022insight}.
Several researchers have reported the use of different photocatalysts since the development of an efficient catalyst can strongly improve the CO$_2$ photoreduction efficiency and selectivity\cite{indrakanti2009photoinduced, chang2016co, fu2020product, li2019cocatalysts, shi2017recent, henderson2011surface}. Among several semiconductors, titanium dioxide (TiO$_2$) has been commonly employed due to its exceptional CO$_2$ photoreduction activity\cite{dominguez2022critical, barrocas2022photocatalytic, wang2022review, li2021recent, zhang2022layered, hossen2022recent, habisreutinger2013photocatalytic}. Recently, CaTiO$_3$ has attracted interest as alternative semiconductor photocatalyst\cite{passi2021review, qiu2024strategies, soltani2021effect, zhang2010electronic} owing to its low cost, great chemical stability, no toxicity and resistance to photo corrosion. Moreover, it can be easily prepared by many methods, being the hydrothermal and solvothermal processes the most popular ones\cite{manjunath2016studies, dong2015simple}. The CaTiO$_3$ perovskite is a n-type chemically stable semiconductor with a wide bandgap of 3–3.5 eV\cite{oliveira2016influence} which allows the absorption of the UV light. However, the latter represents only a small fraction of the solar spectrum (just ~5\% of solar energy) and, along with the high recombination rate of electron-hole pairs, it can severely limit the application of this material\cite{yang2014photocatalytic}. In contrast, it is well known that CaTiO$_3$ has good adsorption features as Ca is a good material to absorb CO$_2$. Indeed, some work has been dedicated on pure CaTiO$_3$ as substrate for the photoreduction of CO$_2$. Kwak et al.\cite{kwak2015photocatalytic} were the first that experimentally investigated the CO$_2$ photocatalytic activity of Ca$_x$Ti$_y$O$_3$ (x = 1.50, 1.28, 1.00 and 0.85; y = 0.75, 0.85, 1.00 and 1.28) with H$_2$O. The observed Ca$_x$Ti$_y$O$_3$ particles were all orthorhombic in shape and the final byproduct detected by gas chromatography was only CH$_4$. The best photocatalytic performance (methane formation rate of $\sim$17 $\mu$mol/$g_{catal}$ for 7 h) was achieved with Ca$_{1.00}$Ti$_{1.00}$O$_3$. Im et al.\cite{im2017effect} used synthesized CaTiO$_3$ with various Ca:Ti ratios coated onto basalt fibers as substrate. Gas chromatography analysis detected a formation rate of $\sim$3.7 and $\sim$11.4 $\mu$mol/g in 8 h relative to CH$_4$ and CO for CaTiO$_3$\@BF , respectively, which increase to a maximum of $\sim$17.8 and $\sim$73.1 $\mu$mol/g for CaTiO$_3$(1.5:1)\@BF. As suggested by the Authors, the latter catalytic sample has the highest amount of TiO$_2$ and oxygen vacancies along with CaTiO$_3$ on basalt fibers among all synthesized samples: all the excited electrons from irradiation moves to oxygen vacancy sites in the rutile TiO$_2$ where CO and CH$_4$ are preferably produced from CO$_2$, which could explain the best performance achieved. 
Many computational studies have been reported to complement experimental works on TiO$_2$-based catalyst by providing atomistic insights underlying the mechanism of CO$_2$ photoreduction, which mainly cover adsorption, activation and reaction pathways occurring on the surface\cite{alli2023photocatalysts, hussain2022theoretical, wu2022dft, yin2016effect, ji2016theoretical, yin2015co2}. In contrast, no similar investigation has been addressed yet on pure CaTiO$_3$ surface, neither CO$_2$ adsorption studies nor early reduction reactions, to the best of our knowledge.
In this work, we report on the density functional theory (DFT) investigation of CO$_2$ photoreduction reactions on pure Ti-terminated CaTiO$_3$ (100) surface for the first time with the aim to elucidate the overall catalytic pathway that leads to the formation of HCOOH, CH$_3$OH and the experimentally detected CO and CH$_4$ byproducts. First, we investigate the adsorption of CO$_2$ and its related intermediates, and then analyze the energetics (i.e., activation energy barriers) of likely and rate-limiting reaction steps.

\section{2. Computational model and methods}
The first-principles method based on DFT was used in the simulations using the Quantum ESPRESSO (QE) code (version 7.3)~\cite{giannozzi2020quantum, giannozzi2017advanced, giannozzi2009quantum}. The generalized gradient approximation (GGA) in the Perdew-Burke-Ernzerhof (PBE) flavor was used ad exchange-correlation (XC) functional~\cite{perdew1996generalized}, along with ultrasoft (US) pseudopotentials. To account van der Waals (vdW) interactions between molecules and the crystal surface the vdW correction terms (DFT-D3) by Grimme et al.\cite{grimme2010consistent} were used. In the DFT-D3 scheme an analytical pair-wise atomic potential is numerically added at the end of each self-consistent energy minimization cycle to the energy values obtained from the Kohn-Sham potential energy functional; the additional computational cost is almost negligible as opposed to vdW-enabled XC functionals (vdW-DF) which directly include the vdW forces in the functional, thus the latter must be evaluated during every self-consistent field cycle. \\
The CaTiO$_3$ (100) surface can have two different surface terminations, namely CaO and TiO$_2$. In this work, the TiO$_2$-terminated CaTiO$_3$ surface with the slab method in the periodic supercell approach was used to study the adsorption and reduction reactions of CO$_2$. To mimic the crystal surface, a lateral 2x2 supercell containing a symmetric five-layer slab (16 Ca atoms, 24 Ti atoms and 64 O atoms) with a 15 \r{A} thick vacuum region was employed to avoid spurious interactions between periodic slab images. The slab was built from the PBE-optimized orthorhombic CaTiO$_3$ primitive cell (Pbnm space group) with lattice constants a = 5.394 \r{A}, b = 5.497 \r{A}, c = 7.679 \r{A}, which are less than 1\% larger than the experimental values of a = 5.372 \r{A}, b = 5.463 \r{A}, c = 7.636 \r{A}. The cutoff energies of 45 and 450 Ry were used for the orbitals and the charge density, respectively, and the $\Gamma$ point was sufficient for the Brillouin zone sampling. The geometries and ground state energies of adsorbed molecules on the crystal surface have been thus investigated in the low-coverage regime (one molecule for eight surface Ti atoms). The molecule was positioned only on one side of the slab. During geometrical optimization the geometry of the bottom two layers was set fixed, whereas the remaining layers with the adsorbed molecule were allowed to relax in all cases until the total energy and forces were less than 10$^{-4}$ Ry (0.0014 eV) and 10$^{-3}$ Ry/Bohr (0.026 eV/\r{A}), respectively. The slab thickness and the supercell lateral dimension were chosen large enough to sufficiently mimic the bulk phase and to avoid spurious interactions among periodic images since periodic boundary conditions were applied. This was checked by looking at the convergence of the adsorption energy, $E_{ads}$, as a function of the supercell periodicity and the number of layers (see Supplementary Information, S1). The adsorption energy $E_{ads}$ is defined as
\[E_{ads} = E_{tot}(molecule+slab) - [E_{tot}(molecule)+E_{tot}(slab)]\]
where $E_{ads}(molecule+slab)$, $E_{ads}(molecule)$ and $E_{ads}(slab)$ are the ground state total energies of the adsorbate-surface system, the isolated molecule in the gas phase and the clean surface, respectively. The entity of the adsorption energy can provide a measure whether a molecule is physisorbed or chemisorbed. In this work, when a negative value was computed a stable adsorption state was obtained, with the most stable structures corresponding to the most negative values. \\
Since GGA suffers from the self-interaction error (SIE) which typically leads to erroneous overdelocalization of strongly localized electrons of d type as in perovskites, the Hubbard correction\cite{anisimov1991band, anisimov1997first} was initially accounted in this work to assess its influence to the description of CaTiO$_3$ by looking at $E_{ads}$ of various CO2 configurations as the target property. The effective Hubbard parameter U of 4.61 eV for the 3d Ti hybrid orbitals was computed by density functional perturbation theory implemented in the hp.x code\cite{timrov2022hp} of QE, which corresponds to a calculated band gap of 3.00 eV (= 2.39 eV without Hubbard) that is consistent to experimental values in the range 3.0-3.5 eV\cite{zhang2010electronic}. In contrast, lattice constants of the CaTiO$_3$ primitive cell and $E_{ads}$ values were found unaltered with respect to the case of U = 0 eV (see Supplementary Information). Hence, all the results presented thereafter were calculated without accounting the Hubbard parameter. \\
The transition state (TS) configuration and the minimum energy path (MEP) of chemical reactions were computed using the climbing image nudged elastic band (CI-NEB) method\cite{henkelman2000climbing} using a convergence force threshold of 0.05 eV/\r{A} for optimization. The initial and final states were deduced from single geometry optimization calculations and a sufficient number of intermediate images was chosen to guarantee an inter-image distance below 1.5 \r{A}.

\section{3. Results and discussion}
Experimental investigations have reported on the detection of carbon monoxide (CO) and methane (CH$_4$) as value-added byproducts of the CO$_2$ photoreduction using the CaTiO$_3$ catalyst\cite{kwak2015photocatalytic, im2017effect}. Even though other plausible chemicals such as methanol (CH$_3$OH) and formic acid (HCOOH) have not been detected, the overall reaction mechanisms leading to the formation of the all above-mentioned compounds were theoretically studied in this work.

\subsection{3.1 Adsorption of CO$_2$}
\begin{figure}[h]
       \includegraphics[width=1.0\linewidth]{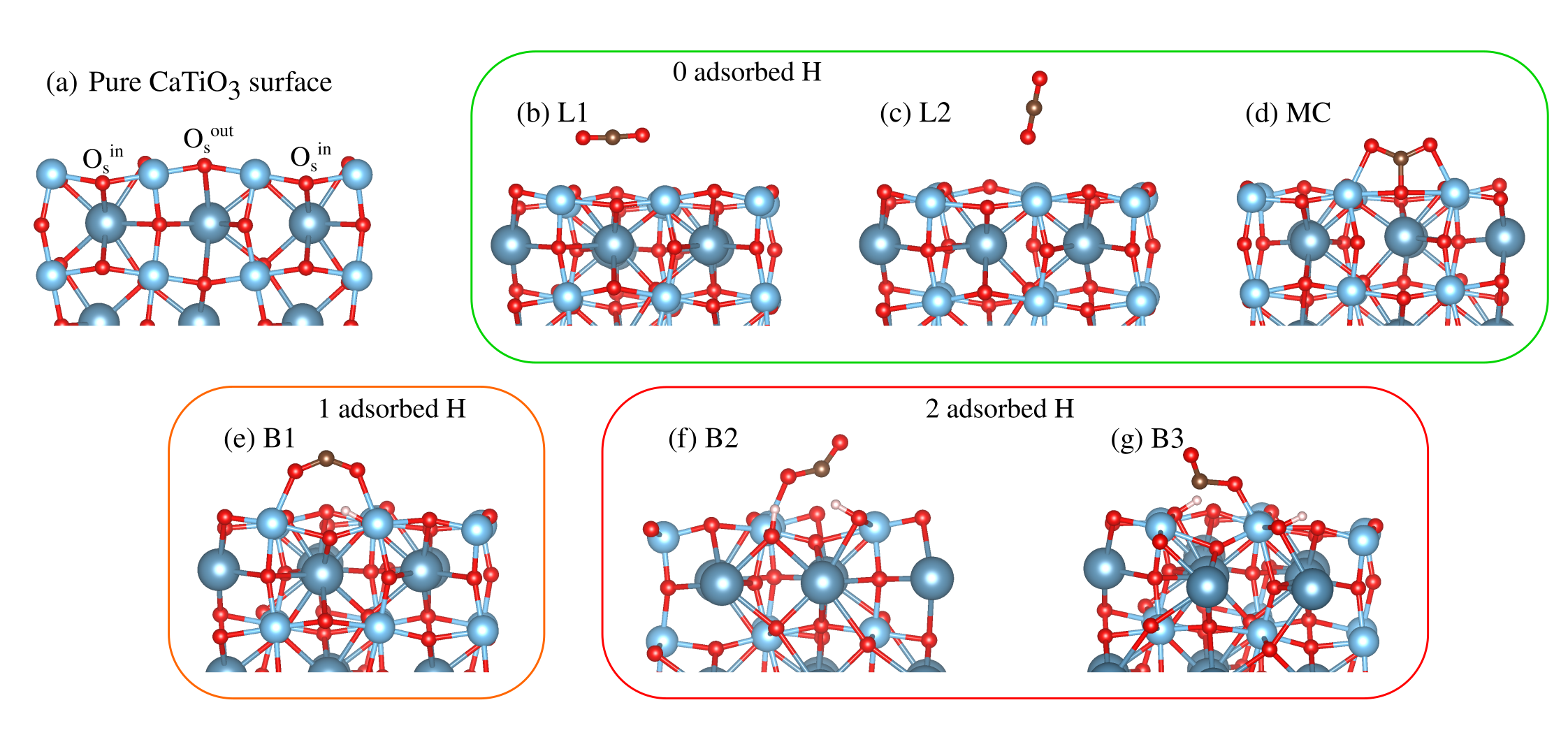}
       \caption{Adsorption configuration of CO$_2$ on TiO$2$-terminated CaTiO$_3$ (100) surface: (a) pure CaTiO$_3$ surface; (b) horizontal linear adsorption 'L1'; (c) horizontal linear adsorption 'L2'; (d) Bridged Carbonate 'BC'; (e), (f) and (g) Bent adsorption 'B1', 'B2' and 'B3'}
       \label{fig:adsorption_co2}
\end{figure}

The adsorption and activation of CO$_2$ on the TiO$_2$-terminated CaTiO$_3$ (100) surface were first investigated as early steps of the photoreduction process. To find different energetically stable adsorbed configurations on the crystal surface, the CO$_2$ degrees of freedom have been progressively varied followed by geometry optimization to eventually achieve new minimum energy structures. This investigation was extended to identify further configurations depending on the adsorption site, i.e. on top of surface Ti and O atoms. Moreover, the surface O atom, indicated as $O_s$ thereafter, can be differentiated by the way it points inwards or outwards to the crystal bulk, as labeled in Fig.~\ref{fig:adsorption_co2}.a by $O_s^{in}$ and $O_s^{out}$, respectively. The activation of CO$_2$ (or any CO$_2$-derived intermediate) by charge transfer of the photoexcited electron to the molecule, as
\[CO_2 + e- \rightarrow CO_2^-\]
was simulated by introducing H atoms adsorbed on top of $O_s$. Indeed, the artificial photoexcited electron available for the charge transfer from the surface to the molecule is drawn from the adsorbed H atom: the latter becomes a proton with apparent charge state +1 when its electron redistributes over Ti atoms to populate the bottom of the CaTiO$_3$ conduction bands. Calculations of atomic spin moments of the activated CO$_2^-$ were conducted to deduce how the photoexcited electron apparently redistributes in the molecule. Indeed, a nonzero atomic spin moment is descriptive of the photoexcited electron localization on that atom. Moreover, charge transfer was also qualitatively demonstrated by removing adsorbed H atoms from activated CO$_2^-$ systems followed by geometry optimization: if a new configuration of inactivated CO$_2$ is reached, thus it means that such H atoms were essential to stabilize the radical anion CO$_2^-$. Chemical states in which charge transfer to molecule occurs are labeled by an asterisk thereafter. \\
\begin{table}[!htp]
    \centering
    \medskip
    \begin{tabular}{p{1cm}cccccccc}
        \toprule
        Conf. & $E_{ads} [eV]$ & C-O$_1$ [\AA] & C-O$_2$ [\AA] & O$_1$-C-O$_2$ & C-O$_s$ [\AA] & Ti-O$_1$ [\AA] & Ti-O$_2$ [\AA] & Ti-C [\AA]\\
        \midrule
        BC & -1.27  & 1.27 & 1.28 & 131.5$^{\circ}$ & 1.35 & 2.13 & 2.16 & / \\
        L1 & -0.43  & 1.18 & 1.18 & 176.8$^{\circ}$ & 2.67 & 2.72 & 2.59 & / \\
        L2 & -0.27  & 1.18 & 1.17 & 178.7$^{\circ}$ & / & 2.49 & 4.66 & 3.55 \\
        B1* & +0.09  & 1.25 & 1.24 & 136.9$^{\circ}$ & 2.95 & 2.13 & 2.12 & / \\
        B2* & +0.42  & 1.27 & 1.21 & 139.4$^{\circ}$ & 2.66 & 2.16 & 3.32 & 2.86 \\
        B3* & +0.27  & 1.42 & 1.21 & 120.0$^{\circ}$ & 2.61 & 1.87 & 3.08 & 2.29 \\
        \bottomrule
    \end{tabular}
    \caption{Adsorption energy, CO$_2$ bond lengths and molecular angle of the various adsorption configurations. Activated configurations are labeled with the asterisk}
    \label{tab:geomCO2}
\end{table}
The obtained adsorbed configurations are shown in Fig.~\ref{fig:adsorption_co2}.b-g, while the CO$_2$ adsorption energies and geometrical parameters are summarized in Table ~\ref{tab:geomCO2}. When it physisorbs, the CO$_2$ molecule can lay horizontally or vertically on the crystal surface, as represented by the L1 and L2 structures in Fig.~\ref{fig:adsorption_co2}.b\&c, respectively. The physisorption nature of the interaction is reflected by the low value of the adsorption energy, which does not exceed -0.43 eV, and by the unperturbed geometry of CO$_2$: the C-O double bond still preserves its length of 1.18 Ang along with the molecule linearity (bond angle of 176.8°), as in the gas phase. Moreover, the surface-molecule distance typically exceeds $\sim$2.5 \r{A}. In the Bridged Carbonate (BC) structure, CO$_2$ strongly bonds with the $Ti$-$O_s$-$Ti$ triplet in the lowest-energy bridged-like configuration (see Fig.~\ref{fig:adsorption_co2}.d) as confirmed by its $E_{ads}$ of -1.27 eV. The resulting C-$O_s$ and Ti-O bonds have lengths of 1.35 and 2.13 \r{A} which are consistent with the estimated single bond values of $\sim$1.4\cite{greenwood1997chemistry} and $\sim$2.1\cite{gomez2021structural} \r{A}, respectively. The CO$_2$ bond angle is 131.5°, which also confirms the strong interaction with the surface. The adsorption surface selectivity plays an important role since CO$_2$ structures can preferentially originate only on certain surface sites that mainly depend on the $O_s$ type, i.e. $O_s^{in}$ and $O_s^{out}$. Indeed, BC will more likely be produced when CO$_2$ gets closer to $O_s^{out}$, otherwise linear configurations as L1 and L2 can be observed above O$_s^{in}$.
When adsorbed H atoms are also accounted, three activated structures were obtained and labeled with the letter ‘B’ (that stands for 'Bent') in Fig.~\ref{fig:adsorption_co2}.e-g. The B1* configuration in Fig.~\ref{fig:adsorption_co2}.e is a well-known structure along with BC that is largely reported on theoretical studies of CO$_2$ photoreduction using TiO$_2$ or other perovskites\cite{wu2022dft, yin2016effect, chernyshova2018origin, nitopi2019progress, terranova2020mechanisms}. In our simulation, the molecule arranges in a bridge-like structure above the surface interacting with two neighboring Ti by its two O atoms, while the unpaired photoexcited electron mainly localizes on the C atom because of its calculated spin polarized moment of 0.6, as also occurs on TiO$_2$. The $E_{ads}$ of the B1* state (+0.09 eV) is significantly higher than that of MC (-1.27 eV) as a result of the molecular electronic reorganization that converts the linear CO$_2$ molecule to the bent radical anion CO$_2^-$. The CO$_2^-$ bond angle decreases significantly to 136.9° and C-O elongates to 1.25 \r{A}, the latter indicating that the double bond character is partially lost, as occurs in BC. Due to adsorption selectivity, the B1* state only originates on top of $O_s^{in}$, whereas the C-$O_s^{out}$ interaction is promoted in the other case which then leads to the BC state. The B2* and B3* configurations in Fig.~\ref{fig:adsorption_co2}.f\&g are the last two activated states identified in this work which resemble to the 'Bidentate Carbonate' structure\cite{wu2022dft, henderson2011surface} because of their chair-like disposition. In Bidentate Carbonate species the CO$_2$ molecule interacts with the Ti-$O_s$ pair via both the O and C atoms, respectively. In contrast, only one O of CO$_2^-$ takes part to bonding with the underneath Ti atom (Ti-O distance of 2.16 \r{A}) in B2* as the C-$O_s$ distance is 2.66 \r{A}, while in B3* both C and O interacts with two neighboring Ti atoms (Ti-C and Ti-O distance of 2.29 and 1.87 \r{A}, respectively) resulting in the C-O bond stretching up to 1.42 \r{A}, thus completely losing its double bond character, and different CO$_2^-$ molecular angle of 120.0°. As opposed to the B2* states, the distance of the interacting Ti-Ti pair in B3* shorten by around 0.2 \r{A} due to steric effects with respect to the case of the clean CaTiO$_3$ surface. B2* and B3* are the least stable activated structures with $E_{ads}$ of +0.42 eV and +0.27 eV, respectively. Both states can be formed when the molecule locates on top of $O_s^{in}$ as in B1*, while the BC state is still favored on $O_s^{out}$. By atomic spin moment analysis, the excess electron cloud in CO$_2^-$ almost equally redistributes between the C and the unperturbed O. In contrast to B1* where only one adsorbed H atom was sufficient to stabilize the activated configuration, two H were essential to produce both B2* and B3* structures: just removing one of them followed by geometry optimization, linear CO$_2$ configurations were obtained.

\subsection{3.2 Photoreduction of CO$_2$ to HCOOH and CO}
To correctly predict and thoroughly account the most likely conversion pathways that involve a series of reduction and/or dissociation reactions from CO$_2$ to HCOOH or CO, we have considered different CO$_2$ adsorption configurations as initial reaction states. The Bridged Carbonate (BC) was found as the lowest-energy state ($E_{ads}$ of -1.27 eV), but the hydrogenation on its O requires an expensive energy barrier, $E_b$, higher than 1.5 eV (see Supplementary Information for further details) which would be unfeasible at typical experimental reaction conditions. The poor reactivity of CO$_2$ in BC resides on its strong adsorption to the surface. Although the conversion into the bent radical anion CO$_2^-$ is affected by high energy costs, it is well accepted by the scientific community that activated molecules are more prone to be reduced and, thus, to be the starting points for the CO$_2$ photoconversion. Hence, only the activated ‘B’ structures were accounted in the overall mechanism thereafter. In this work, each reduction reaction was studied by introducing more than one adsorbed H atom in the system: the first is directly involved in catalytic hydrogenation, while the remaining contributes to the charge transfer. \\
\begin{figure}[htp]
       \includegraphics[width=1.0\linewidth]{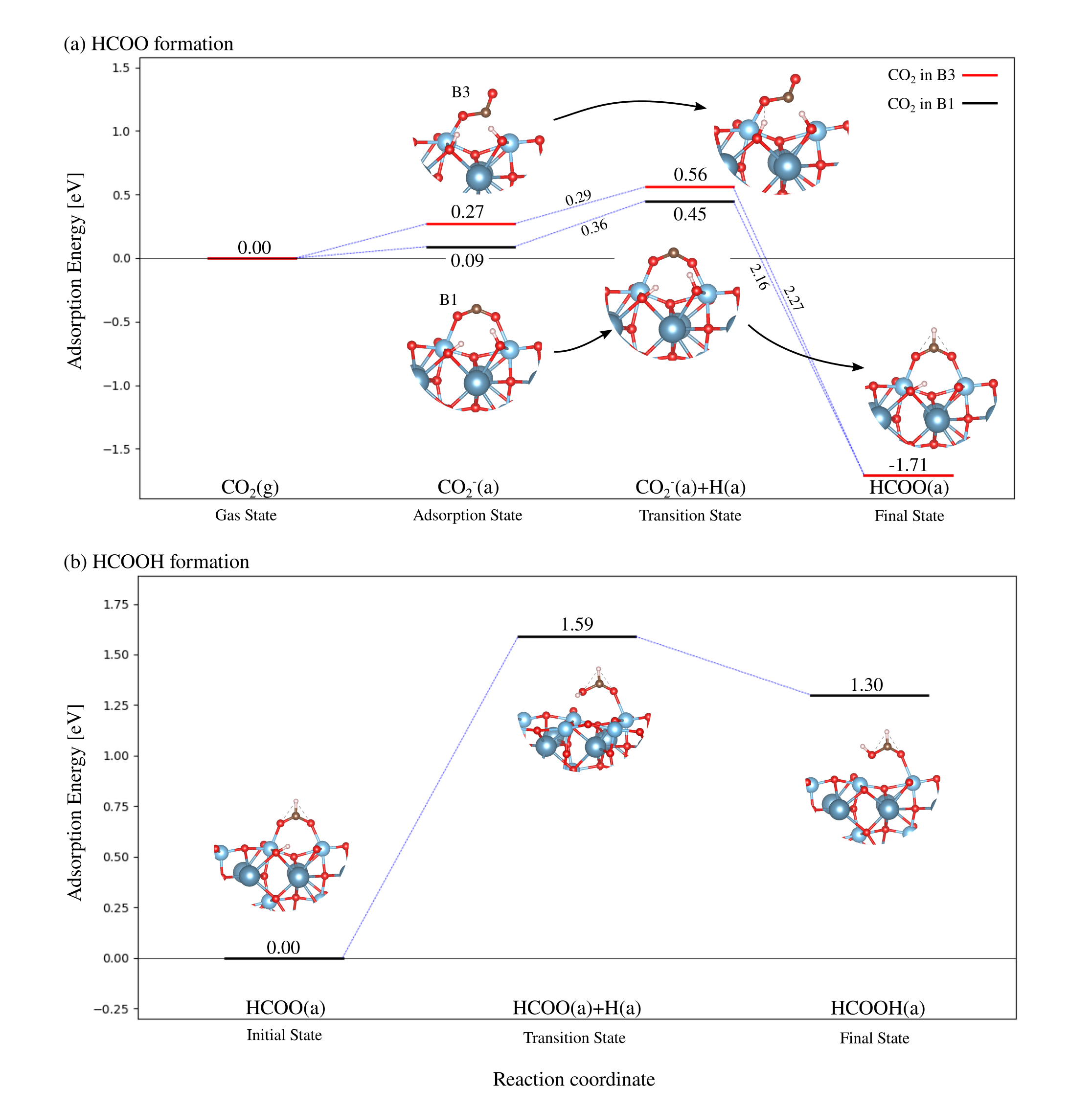}
       \caption{Photoreduction of CO$_2^-$ to HCOOH on TiO$2$-terminated CaTiO$_3$ (100) surface: reaction pathways leading to (a) HCOO from CO2 in B1* and B3*; and (b) HCOOH from HCOO. Activated configurations are labeled with asterisks}
       \label{fig:toHCOOH}
\end{figure}
When the proton approaches the bent anion CO$_2^-$, it can bind to the C or O atom: the formate (HCOO) species is produced in the first case, whereas the carboxylic acid (COOH) in the second case. Afterwards, HCOOH could be finally generated if a second reduction reaction occurs. The minimum energy pathways (MEPs) leading to HCOO from CO$_2^-$ in B1* and B3* are shown in Fig.~\ref{fig:toHCOOH}.a. For the B2* state it was observed that CO$_2^-$ preferably moves first to the more stable B3* state before being hydrogenated on its C since only 0.08 eV are required to overcome the reaction barrier of the B2*-to-B3* conversion. The $E_b$ is 0.36 eV for the B1* case, which slightly lowers to 0.29 eV for B3*. Moreover, the higher chemical stability of HCOO with respect to the 'B' states by around 1.8-2.0 eV hinders the reduction to go backwards. In the B3* mechanism the molecular movement is almost entirely dictated by the proton in the way that it directly locates on top of CO$_2^-$, while the molecule in B3* rotate 180° in order to let its free O atom to bind to the surface and then give formate. It is worth noting that both pathways lead to the same final HCOO configuration which preserves the B1* geometry with C-O distance of 1.27 \r{A} and the Ti-O bond length that slightly changes from 2.13 \r{A} to 2.07 \r{A}. As a result of the C-H bond formation, the molecular angle reduces to 129.0°. Charge transfer to HCOO was not observed as deduced from the following analysis: indeed, by computing and comparing Lowdin charges of equivalent H-reduced and H-unreduced systems one can qualitatively demonstrate if an apparent electron charge increment in the fragment could occur as a result of charge transfer. Proceeding to the second reduction step leading to HCOOH (see Fig.~\ref{fig:toHCOOH}.a) we calculated a prohibitive $E_b$ of 1.59 eV. Even though HCOOH might be formed, the inexpensive reverse reaction ($E_b$ of 0.29 eV) promotes its decomposition to HCOO+H. \\
\begin{figure}[htp]
       \includegraphics[width=1.0\linewidth]{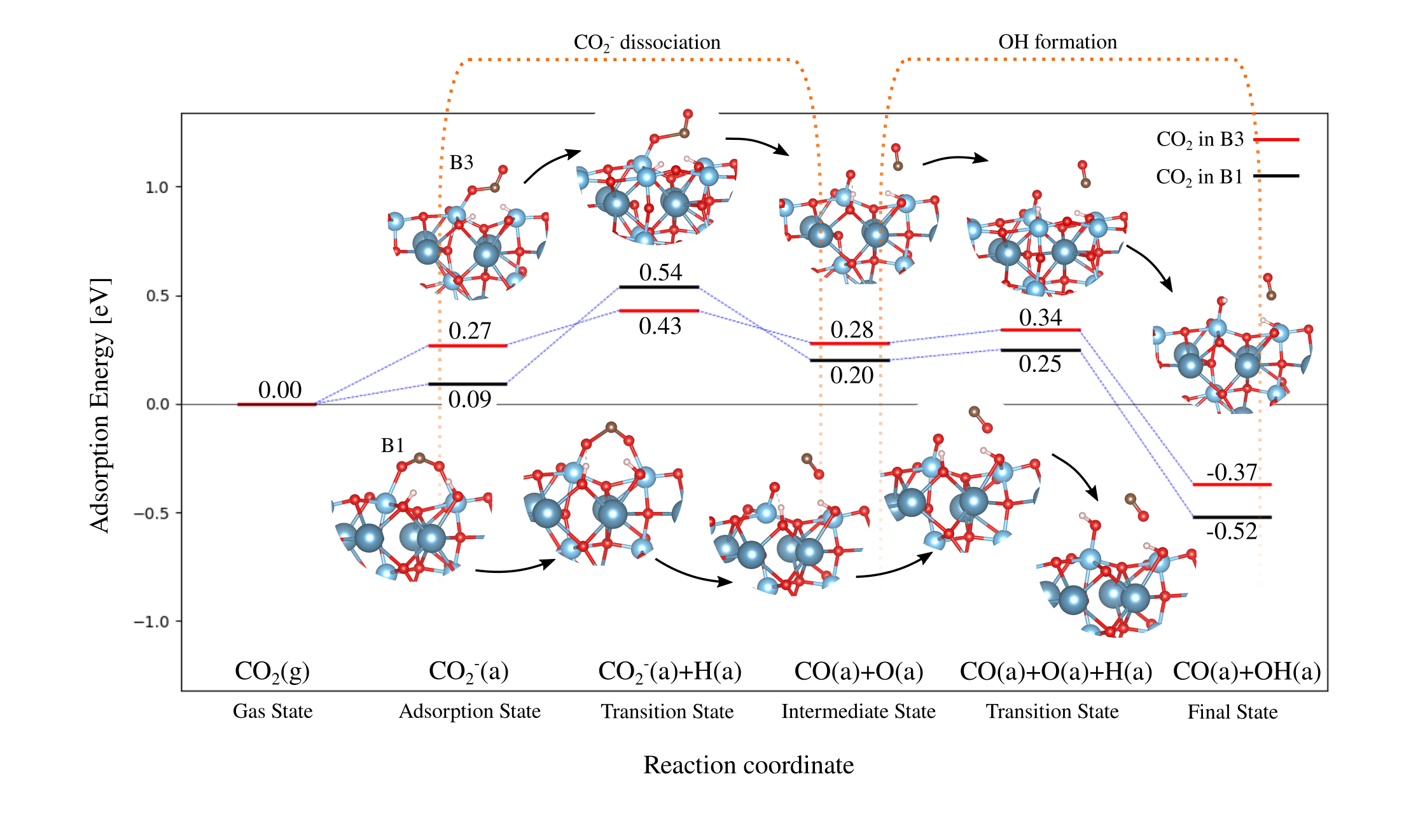}
       \caption{Photoreduction of CO$_2^-$ in B1* and B3* to CO on TiO$2$-terminated CaTiO$_3$ (100) surface. Activated configurations are labeled with asterisks}
       \label{fig:toCO}
\end{figure}
In order to investigate the formation of the COOH intermediate by reduction, we attempted to achieve the final state by progressively approaching the proton to the O of CO$_2^-$ followed by geometry optimization. However, the spontaneous decomposition of CO$_2^-$---H to CO+OH was observed instead of reduction that it would have ideally led to COOH. Thus, we studied the decomposition mechanism of CO$_2^-$ in B1* and B3*. The MEPs are shown in Fig.~\ref{fig:toCO} as combinations of two consecutive elementary steps: (i) first CO$_2^-$ dissociates to give CO+O and (ii) the hydroxyl (OH) species is then produced by the proton attachment to the dissociated O. Only 0.16 eV are required for the B3* decomposition, which increase to 0.45 eV for B1*, although the latter reactant state is energetically more stable than the former by around 0.18 eV. The discrepancy in activation energy values could be attributed to the nature of the surface-molecule interaction. First, it mainly depends on the dissociated C-O bond strength. Since it elongates to 1.42 \r{A} upon adsorption in B3* which has a single bond character, the energy demand to break the C-O bond is lower with respect to the B1* case whose C-O bond (1.24 eV long) partly preserves its double bond character. Second, the Ti-C interaction in B3* enhances the CO production, as opposed to the B1* dissociation in which the as-formed CO adsorbs via its O atom. As it will be explained in the next Subparagraph “\textit{Adsorption of CO}”, CO preferentially physisorbs on top of Ti with its C, as occurs in the intermediate state of the B3* pathway. Following the second elementary step, the OH species is finally produced with an energy cost of 0.06 eV.
Note that the $E_b$ values of the first elementary step and its reverse reaction (i.e. the recombination of CO and O) are almost equal which indicates that a chemical equilibrium could be reached, namely, the concentrations of CO$_2^-$ and decomposed species (CO+O) are comparable. However, the recombination can be hindered by its cheaper competitive process (i.e. the formation of OH species) thus proceeding to the final CO+OH state that is more stable than the intermediate state by around 0.7 eV. The CO+OH recombination was also investigated as it could constitute an alternative pathway to go back to the initial CO$_2^-$ state. However, the mechanism was not observed due to the unreactive adsorbed OH.
From the above analysis, the HCOO and CO species can likely be the most accessible intermediates of the early reduction and decomposition reactions. Afterwards CO could be further reduced to give value-added byproducts like CH$_4$ and H$_3$COH. On the contrary, the high energy cost of the second hydrogenation step of HCOO and the HCOOH proclivity to decompose could explain why formic acid has not been detected in CO$_2$ photoreduction experiments using CaTiO$_3$\cite{im2017effect, kwak2015photocatalytic}.

\subsection{3.3 Adsorption of CO}
\begin{figure}[htp]
       \includegraphics[width=1.0\linewidth]{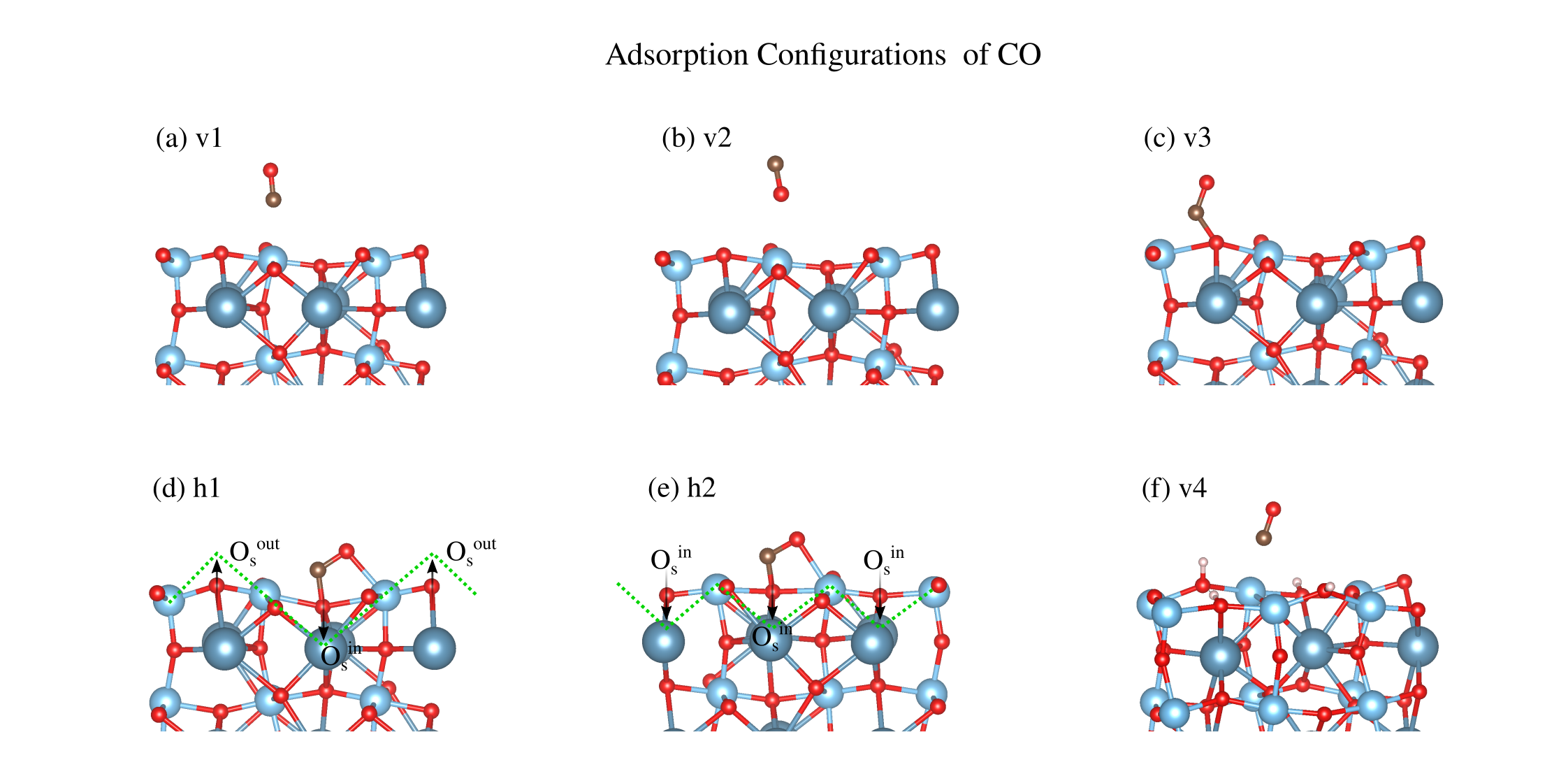}
       \caption{Adsorption configuration of CO on TiO$2$-terminated CaTiO$_3$ (100) surface: (a) v1; (b) v2; (c) v3; (d) h1; (e) h2; v4*. Activated configurations are labeled with asterisks.}
       \label{fig:adsorptionCO}
\end{figure}

Before proceeding to the investigation of the CO photoreduction mechanisms, the CO adsorption study was first addressed in the same spirit of the Subparagraph.3.1. The minimum energy adsorption configurations of CO are all shown in Fig.~\ref{fig:adsorptionCO} and their respective geometrical parameters are summarized in Table ~\ref{tab:geomCO}. 
Physisorption occurs when CO vertically locates on top of Ti and weakly interacts with the surface via its C or O as in the v1 and v2 states, respectively (see Fig.~\ref{fig:adsorptionCO}.a\&b).The Ti-C interaction in v1 is promoted over Ti-O due to the lower $E_{ads}$ of -0.49 eV, which decreases to -0.19 eV for the v2 state. The molecule-surface distance of 2.37 \r{A} in v1 can not be representative of chemisorption since the Ti-C bond in organotitanium compounds is ~2.1\cite{gomez2021structural} \r{A} long. Moreover, the unperturbed C-O bond is 1.14 \r{A} long as in the gas phase. When chemisorbs, CO moves to the v3, h1, h2 or v4* states, as shown in Fig.~\ref{fig:adsorptionCO}.c-f, respectively. Although the low $E_{ads}$ of -0.23 eV, CO in v3 forms a bridging bond with the Ti-$O_s$ pair via its C being the Ti-C and C-$O_s$ distances of 2.11 and 1.38 \r{A} long, respectively. In addition, the CO bond elongates to 1.21 \r{A}. The h1 state is the lowest-energy CO configuration with $E_{ads}$ of -0.61 eV. Although the same C bridging bond in v3 is also found in h1, the higher chemical stability of the latter state originates from the Ti-O covalent bond (distance of 2.16 \r{A}) which further perturbs the C-O bond that now lengthen to 1.26 \r{A}. Due to steric effects of chemisorbed CO, the spatial separation of the interacting Ti-Ti pair increases by around 0.4 \r{A}. CO in h2 shares the same bond's spatial arrangement and similar geometrical parameters of h1, but a lower $E_{ads}$ of -0.09 eV was computed. This energy discrepancy between h1 and h2 can be explained by the adsorption selectivity of $O_s$: the CO adsorption is enhanced when the C atom locates on top of $O_s^{in}$, thus preserving the alternate $O_s^{out}$-$O_s^{in}$-$O_s^{out}$ tuple as in h1 (see dashed green line in Fig.~\ref{fig:adsorptionCO}.d); contrarily, when CO approaches $O_s^{out}$ upon adsorption, the less chemically stable $O_s^{in}$-$O_s^{in}$-$O_s^{in}$ tuple could then be formed as in h2 (see dashed green line in Fig.~\ref{fig:adsorptionCO}.e). As adsorbed H atoms are introduced, charge transfer occurs to CO which moves to the v4* state. By atomic spin moment analysis, the excess electron equally distributes on both atoms. v4* and v1 share the same geometric structure, but the Ti-C interaction in the activated configuration is stronger as evidenced by its reduced bond length of 2.17 \r{A} (with respect to 2.37 \r{A} in v1) and the lower $E_{ads}$ of -0.52 eV.
\begin{table}[!htp]
    \centering
    \medskip
    \begin{tabular}{p{1cm}cccccccc}
        \toprule
        Conf. & $E_{ads} [eV]$ & C-O [\AA] & Ti-C [\AA] & Ti-O [\AA] & C-O$_s$ [\AA] \\
        \midrule
        v1 & -0.49  & 1.14 & 2.37 & 3.49 & / \\
        v2 & -0.19  & 1.14 & 3.69 & 2.61 & / \\
        v3 & -0.23  & 1.21 & 2.11 & 3.27 & 1.38 \\
        h1 & -0.61  & 1.25 & 2.15 & 2.16 & 1.36 \\
        h2 & -0.09  & 1.26 & 2.23 & 2.23 & 1.35 \\
        v4* & -0.52  & 1.16 & 2.17 & 3.32 & / \\
        \bottomrule
    \end{tabular}
    \caption{Adsorption energy, Bond lengths and CO angle of the various adsorption configurations. Activated configurations are labeled with asterisks}
    \label{tab:geomCO}
\end{table}

\subsection{3.4 Photoreduction of CO to H$_3$COH}
On the basis of the CO adsorption analysis in the previous Subparagraph, the v1, v3, v4* and h1 configurations were selected as initial states for the investigation of the H$_3$COH formation. Fig.\ref{fig:toH3COH} reports on the overall reaction mechanism described by four consecutive reduction reactions as a function of the reaction coordinate, along with the energetics of the transition state (TS), the adsorption state (AS) and the final state (FS). All energies are reported with respect to the 0 eV reference state relative to CO in the gas state (GS) far away from the surface. Adsorption processes are assumed barrierless, thus the desorption activation energy equals the binding energy of AS. Moreover, deviations in $E_{ads}$ up to around 0.3 eV were noticed depending on the number and relative position of adsorbed H atoms in the atomic system. As a matter of example, $E_{ads}$ of CO in the h1 state (identified as 'CO-h1' in Fig.\ref{fig:toH3COH}.a) increases by around 0.07 eV for one adsorbed H, which could further increment by around 0.08 eV when a new relative H position on the CaTiO$_3$ surface is accounted. \\

The first CO reduction step (see Fig.\ref{fig:toH3COH}.a) can lead to the HCO or COH intermediates. The adsorbed CO molecule (i.e., in AS) can either desorb and return back to the GS, or undergo reduction to the FS. The outcome of this scenario is driven by the energetic competition between this two events as will be discussed in the follow. For the COH formation, the reduction on O of CO-h1 (red path) requires 1.28 eV since the Ti-O bond must be broken before hydrogenation could occur. As the energy barrier is one order of magnitude higher than that of the competing desorption (0.54 eV), we could suppose that most CO molecules will likely desorb before being reduced to CO-c1.1. The energy barrier lowers to 0.42 eV to give COH-c1.2 from CO-v3 (orange path) as the unsaturated O atom is more prone to be hydrogenated. The order of magnitude of the two competitive events, reduction and desorption of CO-v3, are almost comparable. However, most as-formed COH-c1.2 molecules will preferably decompose and go back to the initial AS since the energetics of the reverse reaction (inexpensive $E_b$ of 0.01 eV) tends to drive the reduction backwards. Very interestingly, the reduction on O of CO-v2 was not observed: if we progressively move the proton to the O of CO-v2 to simulate the COH final state followed by geometry optimization, the CO---H molecule will spontaneously decompose moving to the AS with the proton going back to the adsorbent $O_s$. For the HCO formation, three reaction pathways were investigated. First, the reduction on C of CO-h1 (black path) is promoted by two factor: (i) its low $E_b$ value of 0.58 eV is commensurate to the desorption activation energy of 0.45 eV and (ii) the reaction will likely proceed forward since the as-formed HCO-c1 molecules are energetically more stable than the AS (1.35 eV are required to go backwards). Second, the lower chemical reactivity of the physically-adsorbed CO-v1 hinders the occurrence of the reduction on C (green path) having a prohibitive $E_b$ of 1.31 eV. Third, charge transfer in *CO-v4 supports the first reduction step (yellow path) as $E_b$ halves to 0.63 eV with respect to the equivalent inactivated configuration being now comparable to the activation energy of the competitive desorption event (-0.53 eV). The formation of *HCO-c2.1 is also promoted by the higher stability of the FS over the AS by around 0.57 eV. A more complex situation was observed for the CO-v3 reduction on C (not reported in Fig.\ref{fig:toH3COH}.a). The mechanism can be described by two consecutive elementary steps, namely, (i) the v3-to-v1 conversion which $E_b$ exceeds 1 eV followed by (ii) the hydrogenation that advances through the black path leading to HCO-c1. From the above discussion, HCO-c1 and *HCO-c2.1 could be the most likely intermediates of the early CO reduction as a result of the combination of favored activation energies and highly chemically stable FS. The former intermediate preserves the initial CO-h1 geometric structure, although the C atom no longer interacts with the adjacent Ti (3.12 \r{A} long distance) and the $O_s$ dislocates upwards. Instead, the latter experiences charge transfer and locates on top of Ti with intermolecular distance of 2.24 \r{A}. \\

\begin{figure}
        \centering
       \includegraphics[width=1.0\linewidth]{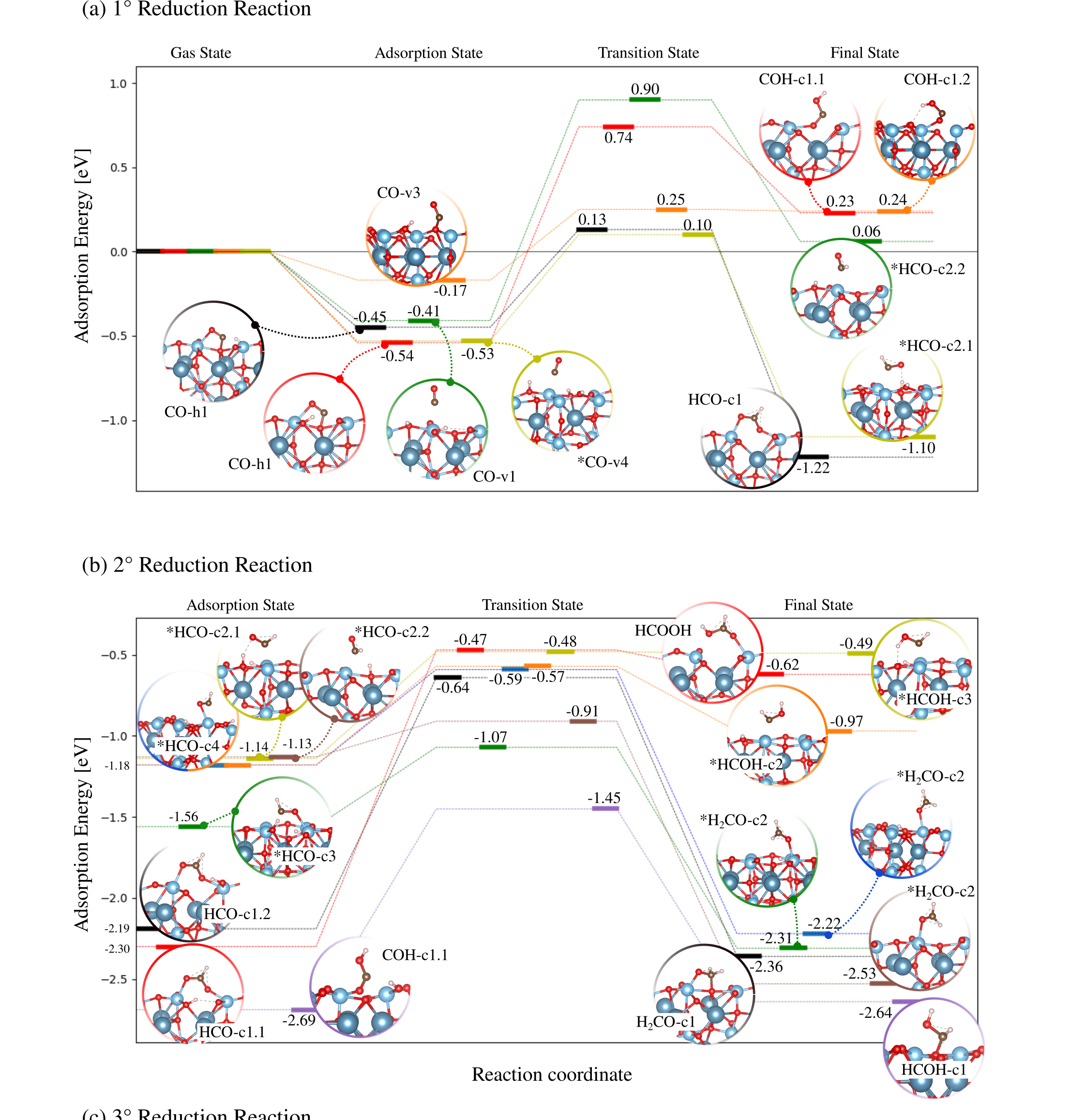}
       \phantomcaption
\end{figure}
\begin{figure}
\ContinuedFloat
       \includegraphics[width=1.0\linewidth]{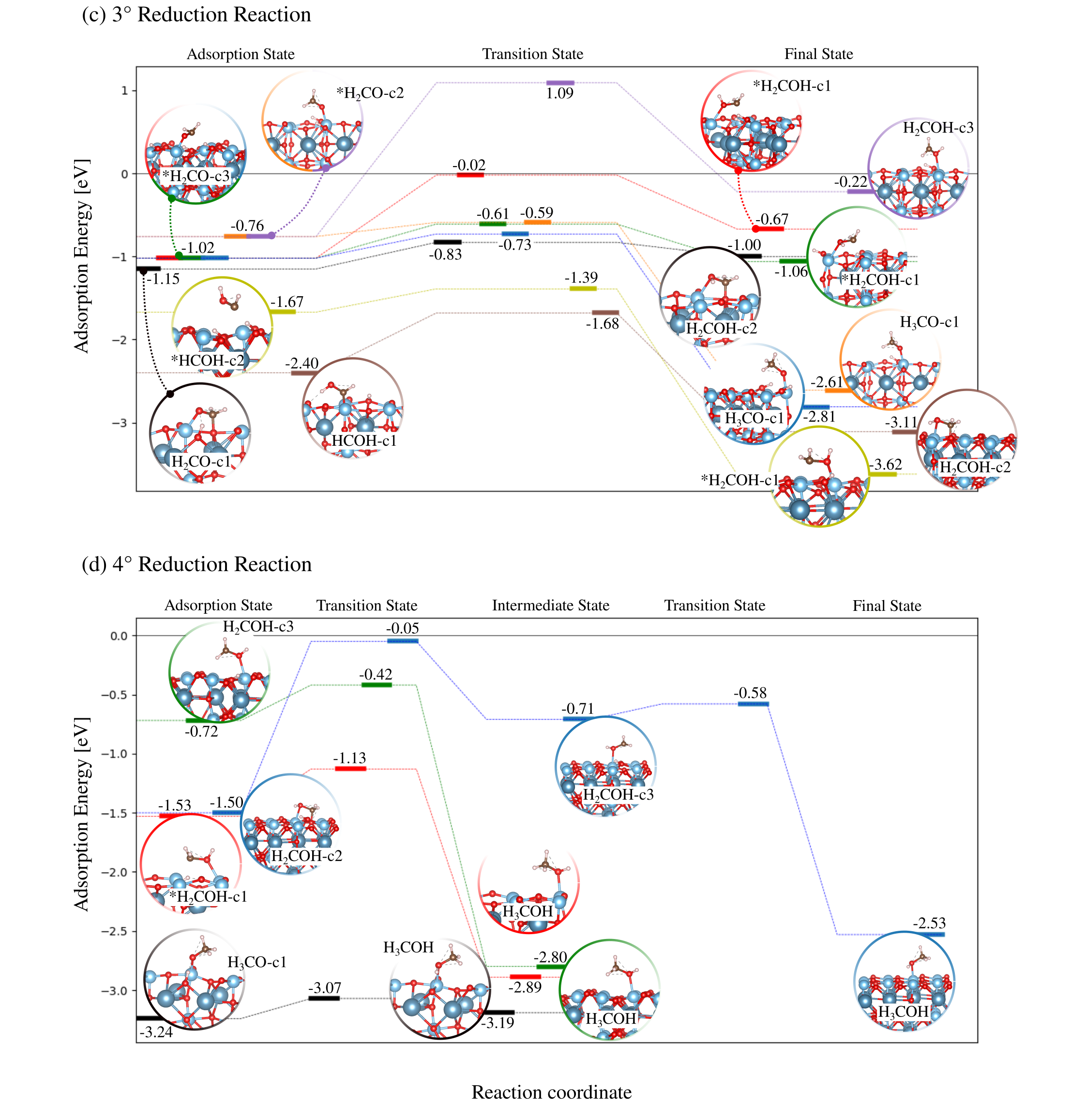}
       \caption{Reaction energy diagram of CO photoreduction to H$_3$COH: (a) first, (b) second, (c) third and (d) fourth reduction reactions}
       \label{fig:toH3COH}
\end{figure}
Fig.\ref{fig:toH3COH}.b shows the second reduction step of HCO leading to H$_2$CO and HCOH. For the sake of completeness, the reduction on C of COH-c1.1 was also accounted even though the production of such intermediate in the early reduction stage is strongly affected by the prohibitive energetics. It is worth noting that the HCO-c1.1 reduction on O has led to HCOOH (red path) instead of HCOH which induces the formation of a $O_s$ defect. Indeed, the HCO fragment withdraws the underneath $O_s$ involved in the C-$O_s$ bond that now becomes part of the molecule. However, the process requires 1.83 eV to give formic acid. The molecule in HCO-c1.1 and HCO-c1.2 has the same geometric structure, but different relative H positions were considered which consequently give distinct $E_{ads}$ of -2.30 and -2.19 eV, respectively. The H$_2$CO formation is first addressed as follows. Among all the HCO configurations, HCO-c1.1 is the lowest-energy state with $E_{ads}$ of $\sim$-2.3 eV. Nevertheless, it can hardly be reduced on C (black path) as 1.55 eV are required to overcome the reaction barrier. To facilitate the hydrogenation, the $O_s$ involved in the C-$O_s$ rearranges moving downwards, which could explain the expensive energy cost. In all remaining HCO configurations, i.e. *HCO-c2, *HCO-c3 and *HCO-c4, charge transfer occurs. *HCO-c3 is the most energetically stable activated structure with $E_{ads}$ of -1.56 eV in which the molecule interacts with a Ti-Ti pair via both its C and O forming Ti-C and Ti-O bonds of 2.33 and 2.23 \r{A} long, respectively. In the $E_{ads}$ range of -1.1$\div$1.2 eV the *HCO-c4 and *HCO-c2 states are found. In *HCO-c4 the C-O bond locates on top of a single Ti atom forming a bridging bond with Ti-C and Ti-O distances of 2.09 and 2.17 \r{A}, respectively, while the surface-molecule interaction in *HCO-c2 is established only by the Ti-C bond (2.24 \r{A} long). The *HCO-c2.1 and *HCO-c2.2 states share the same geometrical structure but they have different molecular orientation: in the latter configuration the O points upwards far away from the surface, while in the former it approaches to the near adsorbed H with which could react and give HCOH. Note that two adsorbed H atoms are required to stabilize *HCO-c3 and *HCO-c4, while only one H is sufficient to simulate charge transfer in *HCO-c2. The lowest energy cost to produce H$_2$CO amounts to 0.22 eV associated to the reduction on C of *HCO-c2.2 (brown path), which increases to 0.49 and 0.59 eV for *HCO-c3 (green path) and *HCO-c4 (blue path), respectively. All three initial states lead to the same H$_2$CO-c2 structure whose FS locates $\sim$1 eV on average below the AS. As a result, the energetics will likely tend to drive the reduction step forwards when activated HCO configurations are involved, thus strongly supporting the H$_2$CO production.
For the HCOH formation, three reduction steps were identified and investigated. A similar energy cost was computed to reduce *HCO-c4 (orange path) and *HCO-c2.1 (yellow path) on the O atom that amounts 0.61 and 0.66 eV, respectively. However, the $E_b$ of the reverse reaction relative to the yellow path is inexpensive (0.01 eV) which supports most of the as-formed *HCOH-c3 to easily dehydrogenate on O going back to the AS, while the orange path is more prone to proceed forwards and to give the more energetically stable *HCOH-c2. The third pathway accounts the reduction on C of COH-c1.1, the lowest-energy structure ($E_{ads}$ of -2.69 eV) among all the investigated configurations of Fig.\ref{fig:toH3COH}.b. in which a strong bridging bond is formed between C and the Ti-$O_s$ pair. The Ti-C and C-$O_s$ bond are 2.18 and 1.27 \r{A} long, respectively. The $E_b$ of this process leading to HCOH-c1 is 1.24 eV. With the information gained in Fig.\ref{fig:toH3COH}.b, we can finally assert that H$_2$CO-c2 can easily be formed from activated *HCO configurations, while only *HCO-c4 could likely give *HCOH-c2 though the reduction can also proceeds backwards. \\
Fig.\ref{fig:toH3COH}.c shows the third reduction stage that leads to the H$_3$CO and H$_2$COH intermediates. As a general trend, HCOH configurations are more energetically stable than H$_2$CO ones. HCOH-c1 has the highest $E_{ads}$ of -2.40 eV which preserves the bond arrangement of COH-c1.1 (see Fig.\ref{fig:toH3COH}.b), although C-$O_s$ now elongates to 1.48 \r{A} with respect to 1.27 \r{A} in COH-c1.1, meaning that the double bond character is completely lost. The $E_{ads}$ increases to -1.67 eV for *HCOH-c2. The surface-molecule interaction is only dictated by the Ti-C bond that is 2.10 \r{A} long, slightly shorter than the equivalent bond (2.24 \r{A} long) in *HCO-c2. Both HCOH structures undergo reduction on C leading to H$_2$COH whose energetics is promoted for *HCOH-c2 (yellow path) demanding only 0.28 eV with respect to 0.72 eV for HCOH-c1 (brown path). The conversion to the FS is further enhanced by its higher chemical stability with respect to AS by around 0.5$\div$1.0 eV. 
Three H$_2$CO configurations were identified among which *H$_2$CO-c2 and *H$_2$CO-c3 are two activated states. H$_2$CO-c1 has higher $E_{ads}$ of -1.15 eV in which the molecule forms C-$O_s$ and Ti-O bonds having lengths of 1.45 and 1.95 \r{A}, respectively. 
The *H$_2$CO-c3 structure with $E_{ads}$ of -1.02 eV preserves the same bond arrangement in *HCO-c3: the C-O bond still interacts with the Ti-Ti pair, but the Ti-O distance now reduces to 1.83 \r{A} in comparison to 2.23 \r{A} of *HCO-c3. The lowest $E_{ads}$ of -0.76 eV was computed for *H$_2$CO-c2 whose surface interaction is regulated by the Ti-O bond. It is worth noting that the same equivalent inactivated structure was also achieved even on the H-unreduced CaTiO$_3$ surface whose surface-molecule distance amounts 2.22 \r{A}. However, when two adsorbed H atoms are also considered, charge transfer occurs leading to *H$_2$CO-c2 with the Ti-O bond that shorten to 2.17 \r{A}. In this work, only the activated configuration was considered. Flattened energy profiles characterized by $E_b$ of 0.32 and 0.41 eV relative to the reduction on O of H$_2$CO-c1 (black path) and *H$_2$CO-c3 (green path), respectively, can promote the H$_2$COH formation. Besides, the FS and AS almost locate on the same energy level which may induce a chemical equilibrium. The red path also accounts the reduction on O of *H$_2$CO-c3 as the green one, but the proton involved in the hydrogenation process comes from a different initial position. However, $E_b$ increases to 1.00 eV which provides a simple and clear example of the expected $E_b$ dependence on the proton position. Aside from the lowest $E_{ads}$, the most expensive energetics is attributed to the purple path relative to H$_2$CO-c2 as the computed $E_b$ is 1.85 eV. The H$_3$CO formation is encouraged by affordable activation energies that do not overcome $\sim$0.3 eV as evidenced by the blue and orange paths relative to *H$_2$CO-c3 and *H$_2$CO-c2. Moreover, its production is further enhanced over the H$_2$COH formation because the FS are more energetically stable than the AS by around 2 eV, thus supporting the reduction step to proceed forwards. The reduction on C of H$_2$CO-c1 was also investigated (not reported in Fig.\ref{fig:toH3COH}.c). The overall mechanism can be described by two consecutive elementary steps, namely, (i) the c1-to-c2 conversion followed by (ii) the catalytic hydrogenation in c2. Step(i) is fundamental to break the C-$O_s$ bond and let the C atom to be hydrogenated. Only 0.53 eV are required to reach the *H$_2$CO-c2 intermediate state, while step(ii) follows the orange path. According to the previous results, the third reduction stage can easily give H$_3$CO as main intermediate since the corresponding energy profiles are characterized by affordable activation energies not higher than 0.3 eV. Besides, only the reduction of *HCOH-c2 could likely lead to a substantial production of H$_2$COH molecules due to the combination of a low $E_b$ of 0.28 eV and a highly energetically stable FS which discourages the reduction reaction to go backwards. \\

Mechanisms of the last reduction step leading to the final H$_3$COH byproducts are summarized in Fig.\ref{fig:toH3COH}.d. H$_3$CO-c1 is the most stable intermediate with $E_{ads}$ of -3.24 eV: the C atom is finally saturated by the O and three H atoms, while the surface interaction is established by the Ti-O bond that is 1.81 \r{A} long. Thus, only the O atom in H$_3$CO-c1 can be hydrogenated. The reaction process proceeds through the black path characterized by a flattened energy profile as the $E_b$ of the forward and backward reaction are smaller and equal to 0.17 and 0.12 eV, respectively. As a result, a chemical equilibrium could be reached. 
Three possible configurations of H$_2$COH were identified. *H$_2$COH-c1 and H$_2$COH-c2 have higher chemical stability with $E_{ads}$ of -1.53 and 1.50 eV, respectively. Here, the molecule interacts with the surface via the entire C-O bond: in *H$_2$COH-c1 the Ti-Ti pair is implicated in the Ti-C and Ti-O bond formation (2.25 and 2.22 \r{A} long, respectively); while, Ti-C is substituted by the C-$O_s$ with bond length of 1.40 \r{A} in H$_2$COH-c2. Adsorption in H$_2$COH-c3 with $E_{ads}$ of -0.72 eV is established by the Ti-O bond that is 2.27 Ang long. The C reduction of *H$_2$COH-c1 (red path) and *H$_2$COH-c3 (green path) requires favorable $E_b$ of 0.40 and 0.30 eV, respectively, which gives H$_3$COH whose FS locates 2 eV lower than the AS. While, the mechanism for H$_2$COH-c2 (blue path) follows two consecutive elementary steps, analogously to the C reduction of H$_2$CO-c1. First, the C-$O_s$ bond must be broken to accommodate the H atom on C leading to the *H$_2$COH-c3 intermediate state by providing 1.45 eV. Second, catalytic hydrogenation can proceeds through the green path.
Once produced the H$_3$COH molecule still locates on the surface adsorbed on top of a Ti atom via the Ti-O bond with length of 2.16 \r{A}. The $E_{ads}$ is -1.01 eV meaning that robust adsorption occurs. \\

In summary, the first reduction stage is the rate-limiting step of the overall mechanism since the corresponding energy profiles are typically distinguished by higher energetics. The COH intermediate yield is crucially inhibited by the expensive energy barriers of the CO reduction process or by the poor chemical stability of COH which easily undergoes dehydrogenation. Instead, HCO could be the main intermediate as the energy cost can decrease to 0.6 eV. The second stage of the CO photoreduction mechanism likely follows the C reduction route leading to H$_2$CO, but low-stability activated HCO configurations must be implied to drastically decrease $E_b$ up to 0.22 eV. One possible investigated pathway could give HCOH from HCO as it requires 0.61 eV, although the reverse reaction is also promoted with the same probability.
The catalytic hydrogenation on C becomes even more favorable in the last two stages, i.e. third and fourth reduction steps, and H$_2$CO and H$_3$CO will constitute the substantial part of intermediates. On the contrary, the reduction on O could reach the chemical equilibrium since the activation energies of the forward and backward reaction are comparable. Even though H$_3$COH can be finally formed, its production yield could be affected by its dehydrogenation on O ($E_b$ 0.12 eV) which competes with the H$_3$COH desorption process that requires 1.01 eV (if barrierless). This could explain why it has not been detected in CO2 photoreduction experiments\cite{kwak2015photocatalytic, im2017effect}.

\subsection{3.5 Photoreduction of CO to CH$_4$}
Contrarily to the H$_3$COH formation mechanism, the CO photoreduction to CH$_4$ includes one additional elementary step dedicated to the O removal from CO-derived intermediates which leads to the production of resulting early hydrocarbon fragments. Then hydrocarbons further undergo catalytic hydrogenation finally leading to CH$_4$. The following issues have been addressed in this work: (i) determine at which stage of the entire mechanism the C-O bond breaking could likely occur and, consequently, (ii) identify the CO-derived fragments that are more prone to lose its O along with the as-formed early hydrocarbons. As a matter of example, the decomposition of HCOH, namely the CO-derived fragment, might produce CH as early hydrocarbon. 
\begin{figure}[htp]
       \includegraphics[width=1.0\linewidth]{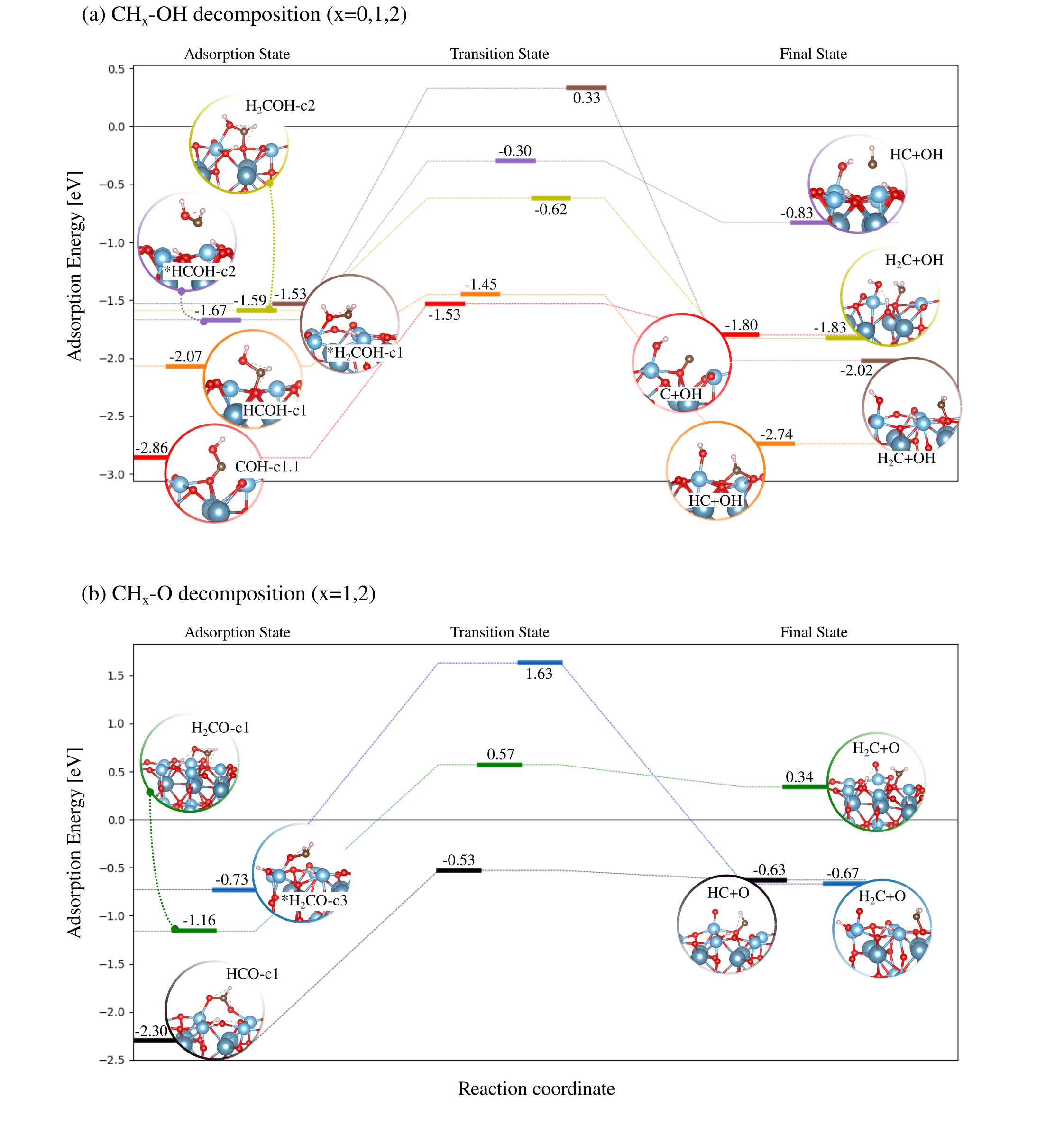}
       \caption{Reaction energy diagram of CO photoreduction to CH$_4$: (a) CH$_x$-OH decomposition (x=0,1,2), (b) CH$_x$-O decomposition (x=0,1,2)}
       \label{fig:toCH4}
\end{figure}
Most of the intermediates identified in the previous Subparagraph have been selected as possible candidates to the decomposition study, as shown in Fig.\ref{fig:toCH4}. Those fragments that do not bear the H on their O atom (CH$_x$O species with x=1,2) are grouped in the panel (b) of Fig.\ref{fig:toCH4} and distinguished by their higher decomposition energy barriers (not lower than 1.73 eV), while the CH$_x$OH species are collected in the panel (a) of Fig.\ref{fig:toCH4}, with x=0,1,2. Although the CO reduction to COH is inhibited, as already discussed, we have evenly accounted the most simple scenario represented by the COH decomposition (red path) that gives the elemental C atom and OH. However, the high $E_b$ value of 1.33 eV for this process is an indication of the low proclivity of the molecule to dissociate. In addition, even though the FS could be reached, the high reactivity of the dissociated elemental C atom would favor its recombination with OH, especially for fast diffusion of the various species. The formation of the simplest hydrocarbon fragment, CH, from the HCOH and HCO intermediates was studied. The HCOH-c1 decomposition (orange path) has the lowest $E_b$ of 0.62 eV among all the investigated pathways. Its occurrence is further enhanced by the higher chemical stability of the initial state ($E_{ads}$ of -2.07 eV) with respect to the other equivalent configuration, namely *HCOH-c2, which locates at -1.67 eV and whose dissociation (purple path) has expensive $E_b$ of 1.37 eV. Moreover, the orange path is discouraged to go backwards since the energy cost to overcome the reverse reaction barrier is higher by around 0.7 eV. The as-formed HC species in the FS forms a strong bridging bond with the $O_s$-$Ti$-$O_s$ tuple. The $E_b$ drastically increases to 1.77 eV when the HCO-c1 decomposition is considered.
Dissociation of H$_2$CO or H$_2$COH can lead to the hydrocarbon fragment CH$_2$. It is worth noting that H$_2$COH-c2 and H$_2$CO-c1 share similar geometric structure in which the O-H bond is only present in the latter configuration, as also observed for the equivalent *H$_2$COH-c1 and *H$_2$CO-c3 states. The H$_2$COH-c2 decomposition (yellow path) has the second lowest $E_b$ of 0.97 eV, while the remaining pathways are characterized by $E_b$ that easily overcome 1.73 eV. As-formed CH$_2$ could then rebind to OH following the yellow path to give the H$_2$COH-c2 molecule in AS, but 1.21 eV are required to overcome the barrier.
The CH$_3$ production was also investigated from H$_3$COH dissociation (not shown in Fig.\ref{fig:toCH4}). The mechanism can be described as a two-step reaction. The first step, i.e. the C-O bond breaking, is the rate-limiting step with $E_{ads}$ of 1.46 eV generating the as-formed metile group in the gas phase. In the second step the CH$_3$ barrierless adsorption occurs. \\
\begin{figure}
        \centering
       \includegraphics[width=1.0\linewidth]{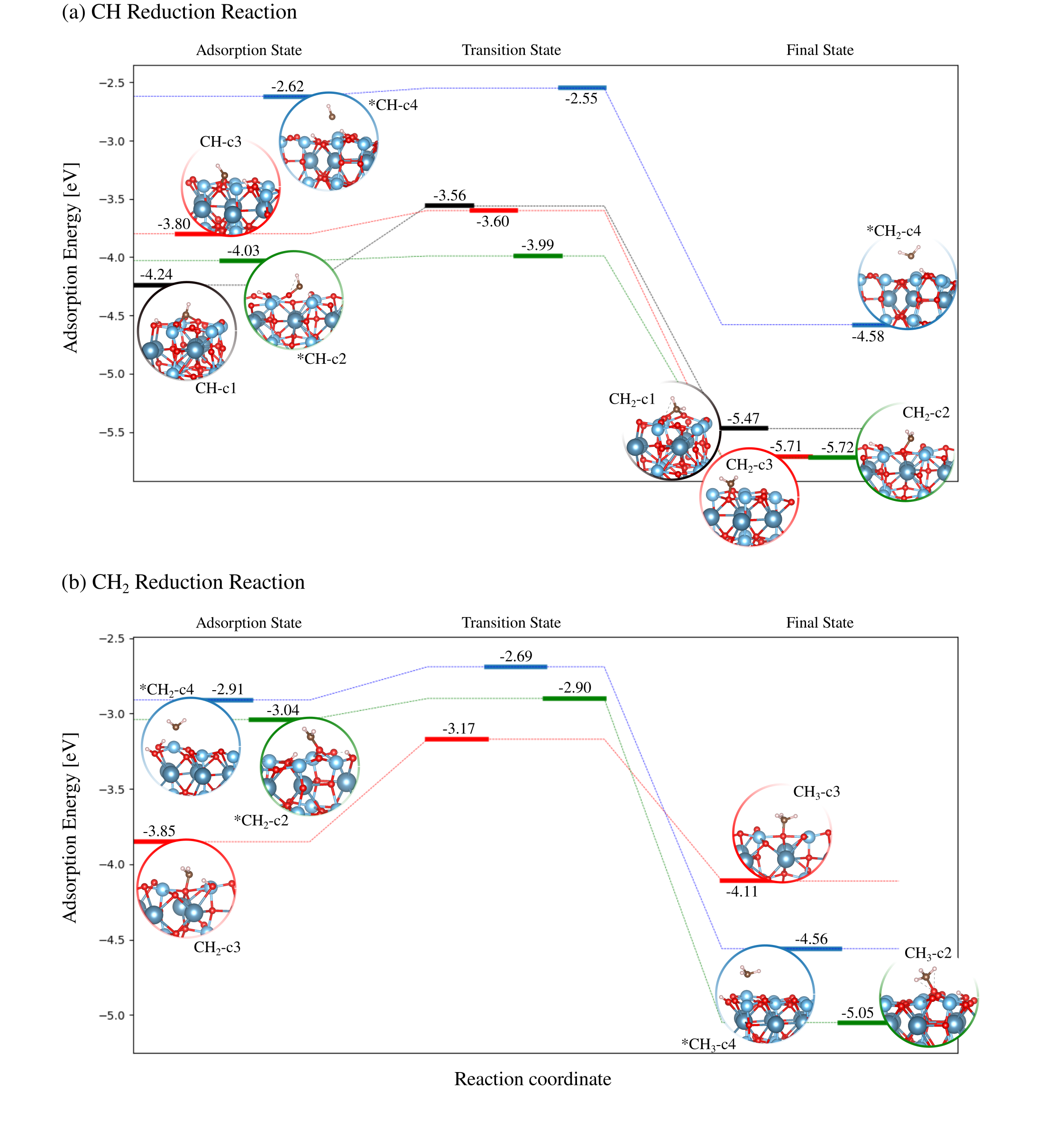}
       \phantomcaption
\end{figure}
\begin{figure}
\ContinuedFloat
       \includegraphics[width=1.0\linewidth]{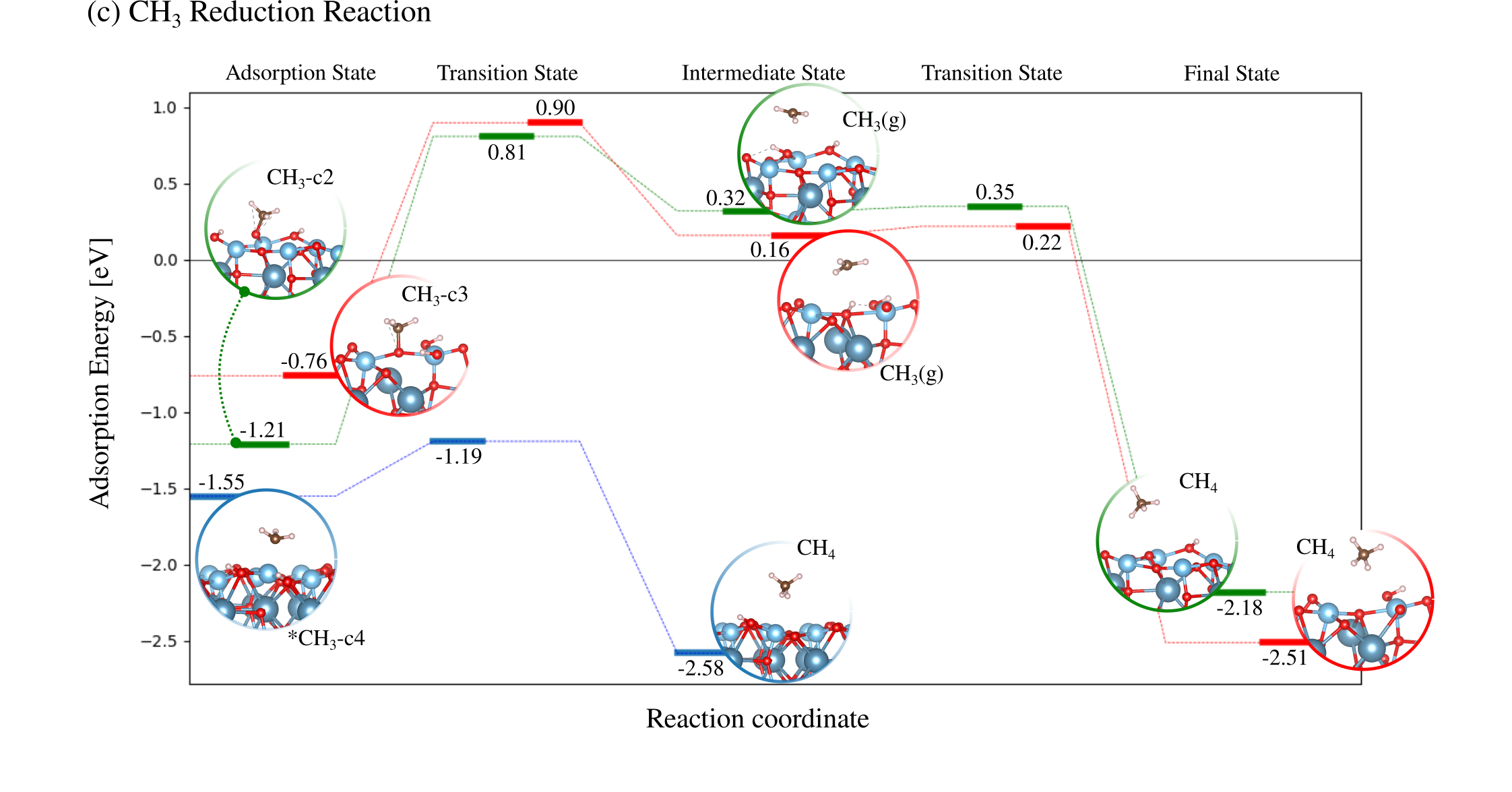}
       \caption{Reaction energy diagram of CH photoreduction to CH$_4$: (a) first, (b) second and (c) third reduction reactions}
       \label{fig:toCH4.2}
\end{figure}

On the basis of the above discussion, CH can likely be the only accessible product of the decomposition stage. Afterwards, three consecutive hydrogenation steps on CH could lead to the final CH$_4$ byproducts, thus following part of the well-known ‘carbene pathway’\cite{habisreutinger2013photocatalytic}. The complete mechanism of the CH photoreduction to CH$_4$ is reported in Fig.\ref{fig:toCH4.2}. Four adsorption configurations ‘c’ can be typically identified for every reduction step since CH$_x$ species (x=1,2,3) can locate (i) between two $O_s$ in a bridge-like disposition (c1, see black paths), (ii) on top of $O_s^{up}$ (c2, see green paths), (iii) on top of $O_s^{down}$ (c3, see red paths) or (iv) on top of Ti (c4, see blue paths). Higher reactivity of hydrocarbon fragments to hydrogenation is observed for activated configurations and for highly unsaturated compounds such as CH. Moreover, the overall energy profiles will tend to drive the reactions forwards as the reverse reactions $E_b$ are prohibitive. The most energetically favorable mechanism follows the overall blue path which involves only activated c4* configurations and whose energy profile has increasing $E_b$ by around 0.1 eV for every hydrogenation step, thus from 0.07 eV for the CH reduction to 0.22 eV for the second hydrogenation which finally extends to 0.36 eV to give CH$_4$. However, activated c4* configurations are characterized by higher $E_{ads}$ values with respect to inactivated ones, apart from *CH$_3$-c4 which has the lowest $E_{ads}$ of -1.55 eV among all the CH$_3$ structures of the third reduction stage. The CH$_x$ species in c4* (x=1,2,3) lay on top of Ti at a distance within the 2.04$\div$2.11 \r{A} range. The CH$_4$ formation route can also easily proceed via the green path as the first two hydrogenation steps occur on activated configurations, namely *CH-c2 and *CH$_2$-c2, requiring only 0.04 and 0.14 eV, respectively. In *CH-c2, C forms a bridging bond with the Ti-$O_s^{up}$ pair with Ti-C and C-$O_s^{up}$ bond lengths of 2.12 and 1.34 \r{A}, while the molecule in *CH$_2$-c2 on top of $O_s^{up}$ (C-$O_s^{up}$ bond length of 1.35 \r{A}) tilts by 56.4°, being the tilting angle defined as the angle between the C-$O_s^{up}$ bond and the surface plane. The last (third) hydrogenation step on inactivated CH$_3$-c2 was observed as a two-step reaction. The CH$_3$-c2 preserves the *CH$_2$-c2 geometric structure with C-$O_s^{up}$ bond that is 1.43 \r{A} long and tilting angle of 65.4°. In order to accommodate the fourth H on its molecular structure, chemisorbed CH$_3$ must first detach from the surface and move to the gas phase, thus assuming the characteristic trigonal planar molecular geometry (see the Intermediate State in Fig.\ref{fig:toCH4.2}.c). However, this process needs to overcome an energy barrier of 2.02 eV. Then, the radical metile in the gas phase requires only 0.03 eV to withdraw the proton from the surface and give CH$_4$. It is evident that the CH photoreduction along the green path can easily proceed until the second hydrogenation step. The energy profile of the CH photoreduction becomes more expensive when the overall red path is accounted. Although the first reduction stage on CH-c3 to CH$_2$-c3 has low $E_b$ of 0.20 eV, the second hydrogenation results less convenient (0.68 eV). In both CH-c3 and CH$_2$-c3 configurations the C interacts with the Ti-$O_s^{down}$ pair with Ti-C and C-$O_s^{down}$ bond lengths of 2.04 and 1.37 \r{A} for the former, or 2.15 and 1.44 \r{A} for the latter. The reduction on CH$_3$-c3 follows the same mechanism observed for CH$_3$-c2 in which a two-step reaction was essential to describe the CH$_4$ formation. The first event again needs a prohibitive energy cost, but now reduces to 1.66 eV. The surface-molecule distance in CH$_3$-c3 is 1.44 \r{A}. Due to the strong binding with the two $O_s$ in CH-c1 as also reflected by its lowest $E_{ads}$ of -4.24 eV, the CH reduction exhibits higher $E_b$ of 0.68 eV. The as-formed CH$_2$-c1 preserves the CH-c1 geometrical structure in which the molecule locates between two $O_s$. However, the hydrogenation on CH$_2$-c1 was not observed since the C is fully saturated by the two $O_s$ and two H atoms. Thus, the C-$O_s$ bond must be broken first to let the proton to attach the CH$_2$ fragment. As a result, the c1-type configurations involved in the mechanism are not optimal for the CH photoreduction process.
Finally, the as-formed CH$_4$ physisorbs with $E_{ads}$ of -0.22 eV at a surface-molecule distance not lower than $\sim$3.2 \r{A} due to its non-polar nature that prevents a strong adsorption on the surface. This could also explain why it has been experimentally detected\cite{kwak2015photocatalytic, im2017effect}, as opposed to H$_3$COH that has higher $E_{ads}$ of -0.93 eV and a good proclivity to dehydrogenate on its O atom when adsorbed.

\section{Conclusions}
We investigated the adsorption and the overall reaction mechanism of the CO$_2$ photoreduction on the Ti-terminated CaTiO$_3$ (100) surface leading to value-added byproducts like HCOOH, CO, CH$_3$OH, CH$_4$. Our study demonstrated that the pathway of the activated CO$_2$ two-step reduction to HCOOH is energetically not feasible although the first catalytic hydrogenation stage can easily produce the HCOO intermediate. This could explain why the HCOOH formation is not typically reported in experimental works. Besides, the conversion of activated CO$_2$ to CO showed low decomposition barriers, in agreement with experimental observations, and it is preferred over the formation of COOH species whose mechanism was not discerned. As one of the most likely abundant intermediates, CO adsorption and photoconversion to CH$_3$OH, CH$_4$ were also addressed. For the CH$_3$OH production, the first reduction step is the rate-limiting step of the entire mechanism, while the energetics becomes more favorable in the last two stages and when activated intermediates are involved. The reaction more likely proceeds through the reduction on the molecule C atom. The O removal in CO-derived intermediates which facilitates the early hydrocarbons formation, then leading to CH$_4$ by subsequent reduction steps, is mainly promoted by the HCOH decomposition giving CH as early hydrocarbon. The combination of charge transfer and intrinsic high reactivity of C$_x$H (x=1,2,3) species strongly support the CH conversion to CH$_4$. Furthermore, CH$_4$ has been experimentally detected as it weakly physisorbs ($E_{ads}$ of -0.22 eV) on the surface, as opposed to CH$_3$OH that has higher $E_{ads}$ of -0.93 eV and a good proclivity to dehydrogenate on its O atom when adsorbed.

\begin{acknowledgement}
O.T. activity was funded by Fondazione Cariplo through the project {\it 2021-0855 SCORE – Solar Energy for Circular CO$_2$ Photoconversion and Chemicals Regeneration}, in the frame of the 2021 call on circular economy. R.R. acknowledges financial support by MCIN/AEI/10.13039/501100011033 under grant PID2020-119777GB-I00, the Severo Ochoa Centres of Excellence Program under grant CEX2019-000917-S, the Generalitat de Catalunya under grant no. and 2021 SGR 01519.  We thank the Centro de Supercomputaci\'on de Galicia (CESGA) for the use of their computational resources.
G.G. acknowledges financial support under the National Recovery and Resilience Plan (NRRP), Mission 4, Component 2, Investment 1.1, Call for tender No. 104 published on 2.2.2022 by the Italian Ministry of University and Research (MUR), funded by the European Union – NextGenerationEU– Project Title 2022LZWKAJ{\_}002 Engineered nano-heterostructures for a new generation of titania photocatalytic films (ENTI) – CUP J53D23003470006 - Grant Assignment Decree No. 966 adopted on 30.06.2023 by the Italian Ministry of University and Research (MUR).
G.G. thanks the European Union - NextGenerationEU under the Italian Ministry of University and Research (MUR) National Innovation Ecosystem grant ECS00000041 - VITALITY for funding and acknowledges Università degli Studi di Perugia and MUR for support within the project Vitality.
\end{acknowledgement}

\begin{suppinfo}
Convergence tests and supplementary reaction steps
\end{suppinfo}


\bibliography{achemso-demo}

\end{document}